\documentstyle[]{mn}
\newif\ifAMStwofonts
\ifoldfss
  \ifCUPmtlplainloaded \else
    \NewTextAlphabet{textbfit} {cmbxti10} {}
    \NewTextAlphabet{textbfss} {cmssbx10} {}
    \NewMathAlphabet{mathbfit} {cmbxti10} {} % for math mode
    \NewMathAlphabet{mathbfss} {cmssbx10} {} %  "   "    "
  \fi
  \ifAMStwofonts
    \ifCUPmtlplainloaded \else
      \NewSymbolFont{upmath} {eurm10}
      \NewSymbolFont{AMSa} {msam10}
      \NewMathSymbol{\upi}     {0}{upmath}{19}
      \NewMathSymbol{\umu}     {0}{upmath}{16}
      \NewMathSymbol{\upartial}{0}{upmath}{40}
      \NewMathSymbol{\leqslant}{3}{AMSa}{36}
      \NewMathSymbol{\geqslant}{3}{AMSa}{3E}

    \fi
  \fi
\fi % End of OFSS

\ifnfssone
  \newmathalphabet{\mathit}
  \addtoversion{normal}{\mathit}{cmr}{m}{it}
  \addtoversion{bold}{\mathit}{cmr}{bx}{it}
  \newmathalphabet{\mathbfit} % math mode version of \textbfit{..}
  \addtoversion{normal}{\mathbfit}{cmr}{bx}{it}
  \addtoversion{bold}{\mathbfit}{cmr}{bx}{it}
  \newmathalphabet{\mathbfss} % math mode version of \textbfss{..}
  \addtoversion{normal}{\mathbfss}{cmss}{bx}{n}
  \addtoversion{bold}{\mathbfss}{cmss}{bx}{n}
  \ifAMStwofonts
    \ifCUPmtlplainloaded \else
      \UseAMStwoboldmath
      \makeatletter
      \new@mathgroup\upmath@group
      \define@mathgroup\mv@normal\upmath@group{eur}{m}{n}
      \define@mathgroup\mv@bold\upmath@group{eur}{b}{n}
      \edef\UPM{\hexnumber\upmath@group}
      \new@mathgroup\amsa@group
      \define@mathgroup\mv@normal\amsa@group{msa}{m}{n}
      \define@mathgroup\mv@bold\amsa@group{msa}{m}{n}
      \edef\AMSa{\hexnumber\amsa@group}
      \makeatother
      \mathchardef\upi="0\UPM19
      \mathchardef\umu="0\UPM16
      \mathchardef\upartial="0\UPM40
      \mathchardef\leqslant="3\AMSa36
      \mathchardef\geqslant="3\AMSa3E
    \fi
  \fi
\fi % End of NFSS release 1

\ifnfsstwo
  \DeclareMathAlphabet{\mathbfit}{OT1}{cmr}{bx}{it}
  \SetMathAlphabet\mathbfit{bold}{OT1}{cmr}{bx}{it}
  \DeclareMathAlphabet{\mathbfss}{OT1}{cmss}{bx}{n}
  \SetMathAlphabet\mathbfss{bold}{OT1}{cmss}{bx}{n}
  \ifAMStwofonts
    \ifCUPmtlplainloaded \else
      \DeclareSymbolFont{UPM}{U}{eur}{m}{n}
      \SetSymbolFont{UPM}{bold}{U}{eur}{b}{n}
      \DeclareSymbolFont{AMSa}{U}{msa}{m}{n}
      \DeclareMathSymbol{\upi}{0}{UPM}{"19}
      \DeclareMathSymbol{\umu}{0}{UPM}{"16}
      \DeclareMathSymbol{\upartial}{0}{UPM}{"40}
      \DeclareMathSymbol{\leqslant}{3}{AMSa}{"36}
      \DeclareMathSymbol{\geqslant}{3}{AMSa}{"3E}
    \fi
  \fi
\fi % End of NFSS release 2

\ifCUPmtlplainloaded \else
  \ifAMStwofonts \else % If no AMS fonts
    \def\upi{\pi}
    \def\umu{\mu}
    \def\upartial{\partial}
  \fi
\fi

\title[Bogus dust screens from well mixed exponential disks in galaxies]
{Bogus dust screens from well mixed exponential disks in galaxies}

\author[B. Elmegreen and D. Block]
  {Bruce~G.~Elmegreen$^1$ and David L. Block$^{2}$\\
  $^1$ IBM Research Division, T.J. Watson Research Center,
        P.O. Box 218, Yorktown Heights, NY 10598\\
  $^2$ Department of Computational and Applied Mathematics,
University of the Witwatersrand, Private Bag 3, WITS 2050, South Africa}

\date{Accepted ...;
      Received ...;
      in original form ...}

\pagerange{\pageref{firstpage}--\pageref{lastpage}}
\pubyear{1998}

\begin{document}

\maketitle

\label{firstpage}

\begin{abstract}
The V-K colours along the minor axes of spiral galaxies typically
change from red to blue with increasing distance, giving the
impression that the near side is systematically screened by dust.
Such a preferred orientation for dust screens is
unlikely.  Here we show that common extinction
from the embedded dust
layer in an exponential disk has the same effect, making the near side
systematically redder as the inclination
increases.  The galaxy NGC 2841 is modelled as an example, where the
V-K profile is profoundly asymmetric and actually step-like
across the centre. We predict that the minor axis emission profile
of the same dust in the far-infrared, at wavelength $\lambda \sim
200\mu$m, will be much more symmetric than the optical
profiles, implying nearly equal column
densities of dust on both sides of the minor axis.\end{abstract}


\section{Introduction}

The recent development of K$'$-band (2.1 $\mu$m ) and K- band
(2.2 $\mu$m) imagery for
nearby galaxy disks led to the unexpected discovery of large V-K$'$
and V-K colour gradients along the minor axes when the inclinations
exceed $\sim 60^\circ$.  The ``Evil Eye'' Galaxy NGC 4826 exhibits a
sudden increase by $\sim$ 1 magnitude from V-K=4 mag. to V-K=5 mag.
in the dust
attenuated `screen' (Block et al. 1994a). Similarly large
V-K jumps are in other highly inclined spirals,
such as NGC 3521 (Panel 188 in Sandage \& Bedke 1994). Explanations
for this effect vary from an intervening dust screen (Witt et al. 1994)
to high latitude scattering from dust
(Block et al.  1996).  The problem with the screen model is that the
apparent
dust is always on the near side of the galaxy, even though we expect
random orientations for galaxies in space.  The problem with the high
latitude dust model is that there is no ubiquitous evidence for such
dust, although extraplanar dust in specific examples has been detected
(eg. Howk \& Savage 1997).

This paper shows how absorption and scattering at V and K bands by
dust embedded in an exponential disk can produce observed colour
gradients which may be remarkably asymmetric or even step-like.
The reason is that the near side of an
exponential disk is brighter at smaller radius
behind the dust in the midplane,
while the far side is brighter at smaller radius in front of the
dust in the midplane. Thus the near side has a higher fraction of the
same total light blocked by dust.

Other models have considered similar effects.  Kodaira \& Ohta (1994)
and Ohta \& Kodaira (1995) measured and modelled V and J band
differential extinction profiles (=$-\ln[I_{far}/I_{near}]$) along the
minor axes in several galaxies, without including scattered light;
they did not discuss colour gradients specifically, although the
existence of such gradients can be inferred from their analyses.
Byun, Freeman \& Kylafis (1994) considered B and I band minor axis
profiles with scattering included, but also did not specifically
address the resulting near-far colour differentials
(although again they can be inferred from their separate B and I profiles).
Kuchinski \& Terndrup (1996)
plotted J-K colour profiles along the minor axes of several galaxies,
but discussed and modelled only the spherical bulge systems, including
scattered light.  Here we determine the line-of-sight V-K colour gradient
in disk+bulge galaxies using radiative transfer with direct
and single-scattered light.

\section{Models}

Stellar and gaseous disks are modelled with exponential
brightness and density profiles in both the radial
and perpendicular directions, following the work of others 
referenced above.
For the stellar
volume emissivity, we use
\begin{equation}
j(r,z)=exp\left(-r-{{|z|}\over{z_s}}\right)+{{\epsilon_b}\over{\left(1-
R^2/R_b^2\right)^{1.5}}},
\end{equation}
and for the dust extinction (scattering + absorption),
\begin{equation}
\kappa_\lambda(r,z)=\kappa_{\lambda,D}(0,0)
exp\left(-r - {{|z|}\over{z_g}}\right)
+{{\kappa_{\lambda,B}(0,0)}\over{\left(1-R^2/R_b^2\right)^{1.5}}} .
\end{equation}
All distances are normalized to the disk scale length; $r$ is the
distance to a point measured parallel to the disk, $R$ is the total
galactocentric
distance, in three dimensions. The volume emissivity is normalized to
unity for each wavelength because we are concerned primarily with
differential (near-far side) colour gradients. The intrinsic stellar
colours do not matter as long as they are the same at each radius on
the near and far side of the disk.  Radial colour gradients in the disk
do not matter for a near-far comparison either.

The equation of radiative transfer gives the intensity $I$ measured
by an observer outside the galaxy at wavelength $\lambda$:
\begin{eqnarray}
\nonumber
I_\lambda(i)=\int_{x_0}^{x_1}dx \left(j[r,z]+j_{scat}[r,z]\right)\\
 \times \;\; exp\left(-\int_{x_0}^{x}\kappa_\lambda[r,z]dx^\prime\right).
\end{eqnarray}
The distance along the line of sight is $x$, increasing away from the
observer; $x_0$ and $x_1$ are the near and far limits of the
integral through the galaxy, taken here to
be the hypotenuse of a triangle with a
maximum total in-plane radial extent of 5 scale lengths and
a maximum total half-thickness
of 3 scale lengths.  The argument of $I_\lambda$, $i$, is the
galaxy inclination.

\begin{figure}
\vspace{5.0in}
\includegraphics{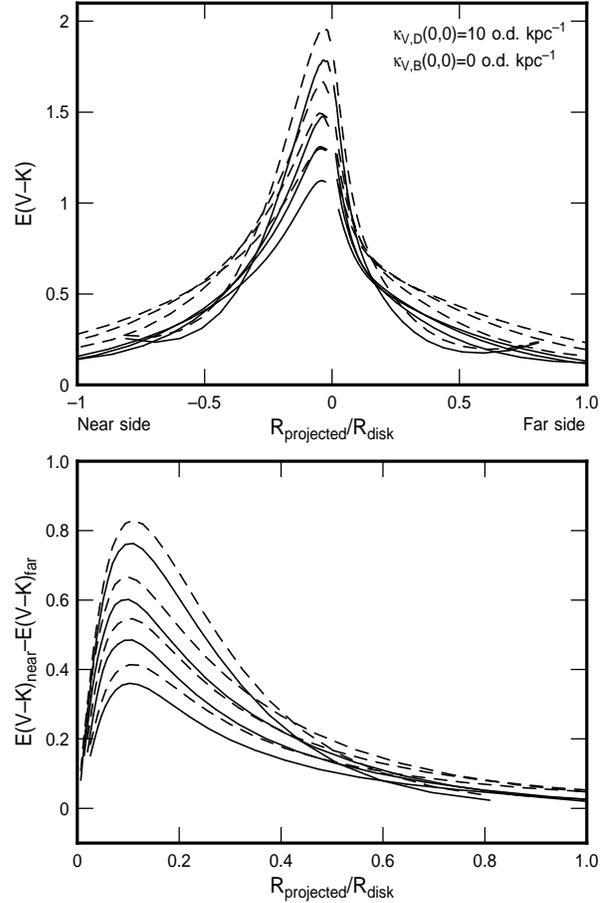}
%run 39
\caption{(top) V-K colour excess profiles along the minor axes of model
galaxies with inclinations of 50$^\circ$, 60$^\circ$,
70$^\circ$, and 80$^\circ$ ($E[V-K]$ increases with inclination).
Solid lines include scattered light.
This case has no dust distributed in the bulge.
(bottom) V-K colour excess differences between the near and far sides
shown as functions of minor axis distance.
}
\label{fig:nobulge}
\end{figure}

The volume emissivity from single scattering is
\begin{eqnarray}
\nonumber
j_{scat}(r,z)=\kappa_\lambda(r,z)A_\lambda
\int_{-\pi}^\pi dl \int_{-\pi/2}^{\pi/2} db \cos b \phi(\theta)\\
\times \int_0^{x_2}j(r,z) d\xi
exp\left(-\int_0^{\xi}\kappa_\lambda[r,z]d\xi^\prime\right).
\end{eqnarray}
where $x_2$ is the maximum distance along a path through the galaxy,
$l$ and $b$ are the galactic longitude and latitude as seen by
an observer at point $x$ in the integral for $I_\lambda$.
$\phi$ is the scattering phase function (Henyey \& Greenstein 1941),
\begin{equation}
\phi(\theta)={{1}\over{4\pi}}{{1-g^2}\over{\left(1+g^2
-2g\cos\theta\right)^{1.5}}}
\end{equation}
and $\theta$ is the scattering angle, given by
\begin{equation}
\cos\theta=-\sin b\cos i + \cos b \sin i \cos l.
\end{equation}

In equation (4), $A_\lambda$ is the albedo at wavelength $\lambda$.
While
most interstellar grains were thought to be substantially smaller than
near-infrared wavelengths (Mathis, Rumple \& Nordsieck
1977), the  upper size
limit of dust grains -- to which the near-infrared albedo is extremely
sensitive -- has been revised (see Kim, Martin \& Hendry 1994).
The first extragalactic determination of the near-infrared
dust albedo was made by Witt et al. (1994), who concluded
that the albedo of dust grains at V and K may be identical.
This high near-infrared albedo is alluded to by a number of diverse
observations (Lehtinen \& Mattila 1996 for the Thumbprint Nebula;
Block 1996 for the Whirlpool Galaxy M51 and its companion;
Pendleton, Tielens \& Werner 1990 and Sellgren, Werner \& Dinerstein
1992 for infrared and classical reflection nebulae).
We choose  values for the optical and near-infrared albedos of dust
grains to both be $A=0.6$ in the V and K passbands.

Furthermore, dust grains are
predominantly forward scattering in the optical (eg. Whittet 1992)
while
near-infrared values of $g$ at K lie between 0.1 and 0.6 (Witt et al.
1996). We choose $g=0.5$ in V and K.
There is only a weak dependence on $g$ as far as albedo
determinations are concerned (Fig 2 in Witt et al. 1996), and our
models with different values of $A$ did not
change much; lower $A$ gave about the same results as slightly higher
extinctions, approaching the results with no scattering,
as shown in the figures by dashed lines.

\begin{figure}
\vspace{5.0in}
\includegraphics{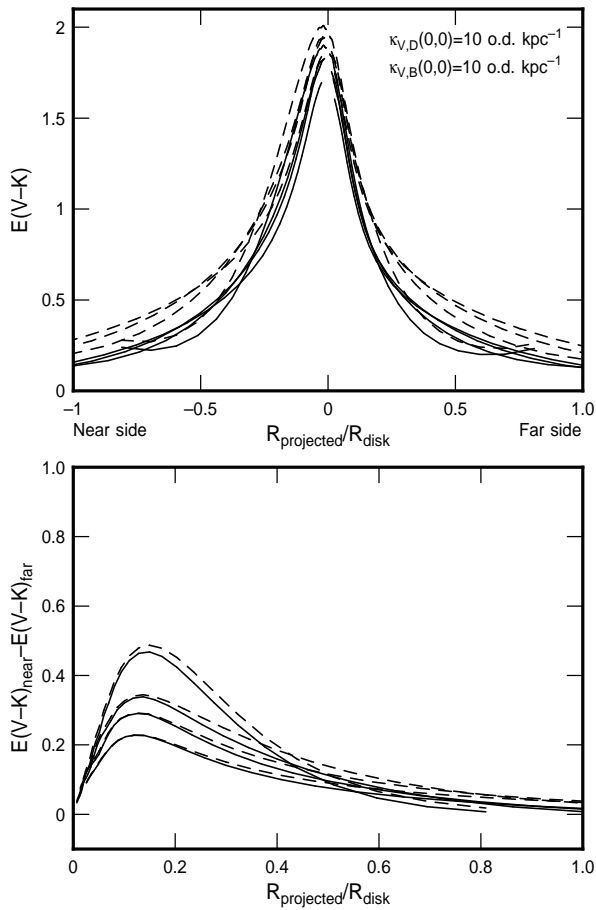}
% run 63
\caption{Same models as in the
previous figure, but with dust distributed throughout
the bulge region, in three-dimensions.}
\label{fig:bulge}
\end{figure}

The intensity $I_\lambda$ was determined from these equations
as a function of distance from the galactic center
for various galaxy inclinations. For each distance and inclination,
the reddening difference between the near and far sides
of the disk was evaluated in magnitudes:
\begin{eqnarray}
E(V-K)_{\rm near}-E(V-K)_{\rm far}=\\
-2.5\log_{10} \left({{I_V}\over{I_K}}\right)_{near}
+2.5\log_{10} \left({{I_V}\over{I_K}}\right)_{far}.
\end{eqnarray}

For most of the models shown here, the perpendicular scale
heights are taken to be $z_g=0.1$ and $z_s=0.2$,
and the bulge scale
length is $R_b=0.1$ (Kodaira, \& Ohta 1994).
The relative volume emissivity of
the bulge is $\epsilon_b\sim0.5\left(L_b/L_d\right)z_s/R_b^3$ (Kodaira
\& Ohta 1994). The ratio of bulge to disk luminosity
is $L_b/L_d=0.1$ at K band
(Byun, Freeman \& Kylafis 1994) and $0.1\times10^{-0.4 \Delta M}$ at $V$ band
for colour difference $\Delta M=0.5$ mag between the bulge and the
disk.
The central extinctions for the disk
are taken to be $\kappa_{V,D}(0,0)=10$ or 15 optical depths (o.d.) kpc$^{-1}$
for the models shown by the figures,
and for the bulge are
$\kappa_{V,B}(0,0)=0, $ or 10 o.d. kpc$^{-1}$
(1 optical depth = $0.4/\log e = 0.92 $ magnitude.)
In all cases, $\kappa_K=0.1\kappa_V$.
The value of $\kappa_{V,D}(0,0)=10$ o.d. ${\rm kpc}^{-1}$
might be appropriate for the Milky Way: it
gives a V-band extinction of 1.25 mag kpc$^{-1}$ at $r=2$ disk
scale lengths, which is about the extinction at the solar radius.
Many other values for these quantities were modelled too, and
some of the results will be discussed for comparison.

Figure \ref{fig:nobulge}(top) shows the
intensity versus distance along the minor axis for a model with
a central disk extinction of 10 o.d. kpc$^{-1}$ in $V$
and no additional extinction in the bulge.
The projected distance along the minor axis, $x\cos i$,
is on the abscissa, in units of the intrinsic disk
scale length; negative values of this distance
represent the near side of the galaxy.
Both the bulge
and disk emissions are included here, but there is no bulge
extinction in this model.
Cases with scattered light are shown by solid
lines, and cases without scattered light are shown by dashed lines.
Inclinations of 50$^\circ$, 60$^\circ$, 70$^\circ$,
amd 80$^\circ$ are plotted; the near-side reddening increases
with inclination.
Any intrinsic disk colour gradient
that might be present in the stars is ignored.

The V-K colour differences between the near and
the far sides are shown in the bottom panel of figure \ref{fig:nobulge}.
There is
a maximum in differential reddening at about 0.2 disk
scale length, where the visual extinction is $\sim8$ o.d. kpc$^{-1}$.
Intrinsic disk colour gradients would not matter for this
diagram, as long as the near and far sides of the disk
have the same intrinsic colours at the same radii.

Figure \ref{fig:bulge} shows
the same model as figure \ref{fig:nobulge}
but with extinction in the bulge. For the assumed central
bulge extinction of
$\kappa_{V,B}(0,0)=10$ o.d. kpc$^{-1}$,
the extinction at a radius of 1 disk scale
length is $10\;{\rm o.d.\;kpc^{-1}}
\left(1+\left[1/0.1\right]^2\right)^{-3/2}=0.01$ mag kpc$^{-1}$,
which is much smaller than the disk extinction at this radius.
Nevertheless, dust in the bulge clearly reddens the far side of the
$V-K$ colour profile, making the V-K colour difference at the
bottom of the figure about a factor of 2 less than in figure
\ref{fig:nobulge}.

\begin{figure}
\vspace{2.8in}
\includegraphics{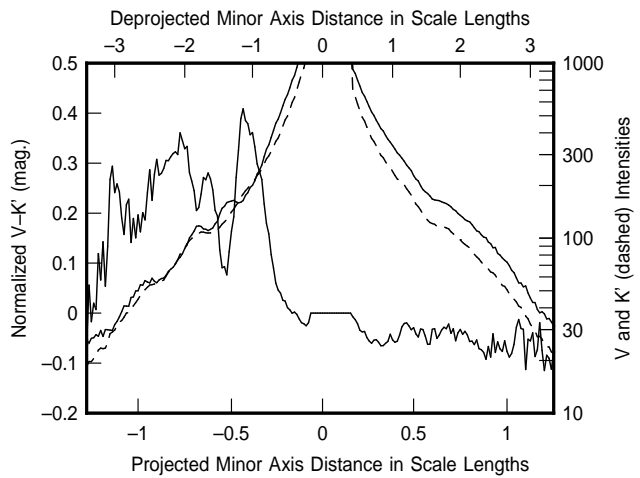}
\caption{Minor axis V-K$^\prime$ profile for the galaxy NGC 2841, showing
redder colours on the near side (left)
than the far side. The axis for the colour difference is
on the left. Also shown are the V and K$^\prime$ intensity scans across
the minor axis, using the axis on the right. The dashed line is
K$^\prime$.}
\label{fig:n2841scan}
\end{figure}

Figure \ref{fig:n2841scan} shows the observed V-K$^\prime$
scan along the minor axis of the inclined flocculent galaxy
NGC 2841 (studied in more detail by Block et al. 1996).
The strong colour gradient between the near (left in the figure)
and far sides is evident, as discussed in our previous paper.
The models in figures \ref{fig:nobulge}
and \ref{fig:bulge} do not reproduce this gradient
well because NGC 2841 has a central hole in molecular gas inside 1
disk scale length (Young \& Scoville 1982), and it has a low HI
column density there too. Thus there is probably a dust hole
inside 1 scale length.
This hole was also evident in a scan of V-K colour
versus distance
along the major axis of NGC 2841 (Block et al. 1996),
which shows a dip toward bluer colour inside 1 disk
scale length. This blueness approaches the true
colour of the inner disk, without extinction.
In addition to the dust hole,
NGC 2841 also has a relatively high extinction perpendicular
to the disk compared to the Milky Way (Block et al. 1996),
so we need to consider
slightly larger $\kappa$ than in the previous figures.

\begin{figure}
\vspace{5.0in}
\includegraphics{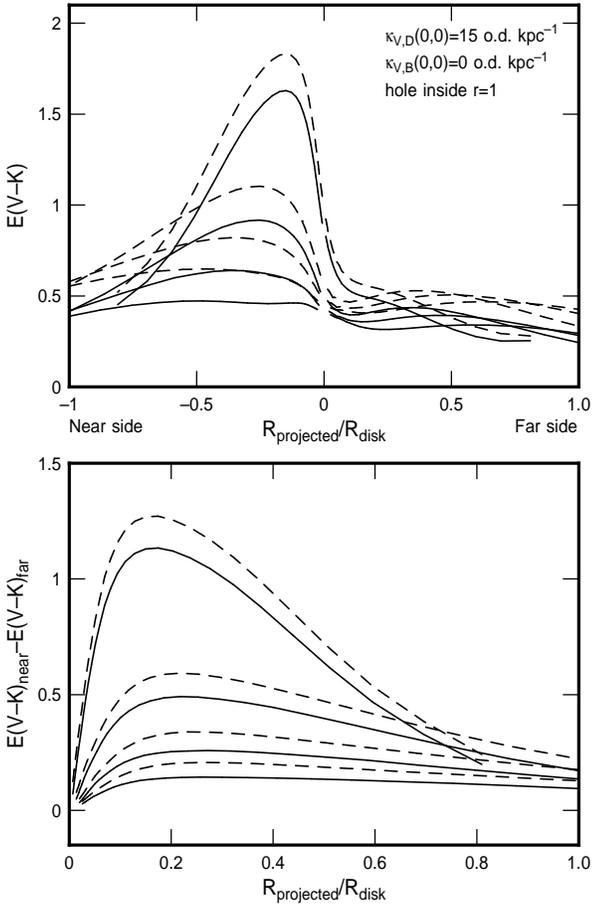}
\caption{Model appropriate for NGC 2841 with four inclinations i
as before. In this
case the inclination $i=70^\circ$, which is the
second from the top, most closely matches the inclination of NGC 2841.}
\label{fig:n2841model}
\end{figure}

Figure \ref{fig:n2841model} shows the minor axis profile from
a model with 1.5 times the extinction
as in figure \ref{fig:nobulge} but
with a central hole made by setting the dust density profile
equal to $re^{-r}$ instead of $e^{-r}$ for radius $r$, measured
in units of the scale length.
This function gives a peak in the gas density at $r=1$ with a
hole inside this and a nearly exponential disk beyond.
The model has the same four inclination angles as in the previous plots,
for general use in other studies; the inclination of NGC 2841 is
$\sim68^\circ$, which is essentially the same as for the third curve
up from the bottom (which is for $70^\circ$). With a hole in the model
and no dust in the bulge, the fit to the data is good, i.e.,
the near side is redder than the far side by $\sim0.4$ mag., and
the far side colour excess distribution is relatively flat.
The small scale variations in the data for NGC 2841 are
presumably from dust lanes, which are not part of the present model
(but see Block et al. 1996
for radiative transfer fits to these dust lanes).

\begin{figure}
\vspace{5.0in}
\includegraphics{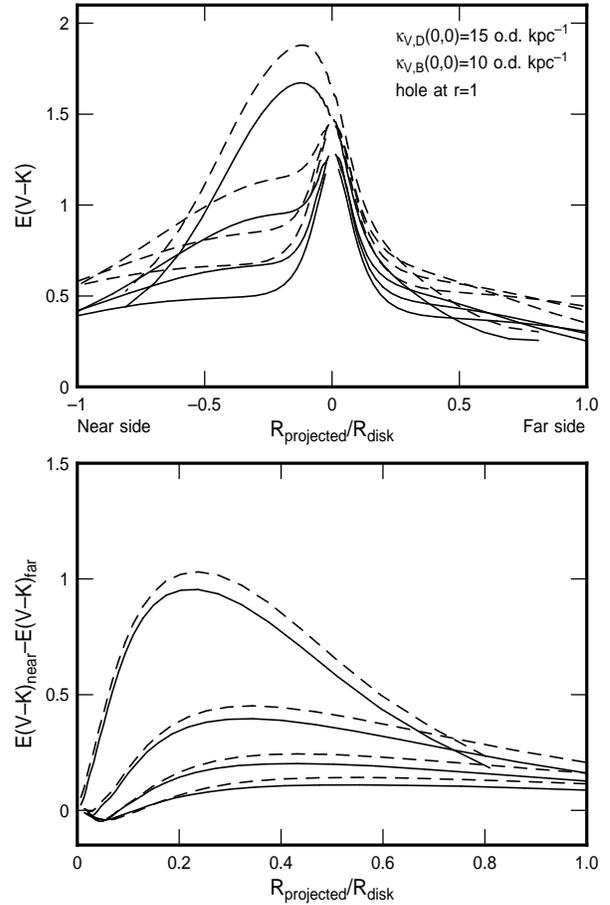}
\caption{Model with a disk dust distribution as in Fig. 4, but with dust
also in the bulge, off the plane. This case is not a good
match for NGC 2841 because the bulge dust reddens the far side of
the disk, and produces too high a colour differential there.}
\label{fig:n2841modelbulge}
\end{figure}

Figure \ref{fig:n2841modelbulge} shows the same disk-hole model as in figure
\ref{fig:n2841model}, but with extinction in the bulge as in
figure \ref{fig:bulge}. This model does not fit the data for NGC 2841 because
the far side is too red with the additional extinction from the bulge,
giving the far-side part of the curve on the top of figure
\ref{fig:n2841modelbulge}
a downward slope, unlike the observations of NGC 2841
and the corresponding curve in figure \ref{fig:n2841model}, where this
far-side E(V-K) scan is flat.
The dusty-bulge model also has an increased reddening in the central
region, which becomes even more prominent for face-on orientations.
Any larger $K_{V,B}(0,0)$ makes this central reddening
unacceptably large even for the inclination of NGC 2841.
Thus the upper limit to the amount of bulge dust in NGC 2841 is
about the value in figure \ref{fig:n2841modelbulge}, which
corresponds to less than 0.01 o.d. kpc$^{-1}$ at one scale length,
or an equivalent hydrogen density of $<0.007$ cm$^{-3}$, using
the standard conversion of dust to gas
(Bohlin, Savage \& Drake 1978).
Evidently our previous suggestion (Block et al. 1996) that the
near-far colour difference in NGC 2841 is the result of bulge scattering
by dust high off the plane is not correct.

\begin{figure}
\vspace{4.6in}
\includegraphics{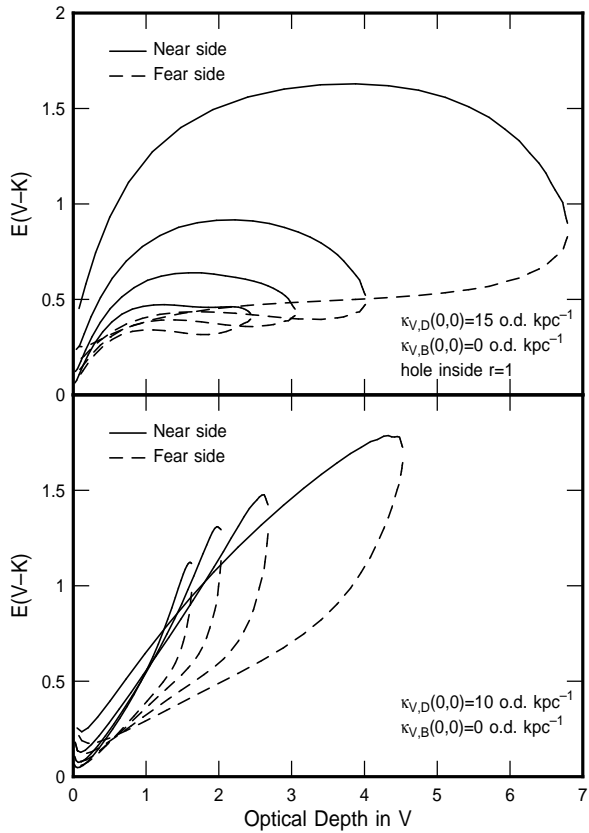}
\caption{Color excess versus optical depth through the disk for the Milky
Way and NGC 2841 models shown previously in figures 1 and 4.
The lack of a simple linear relationship results from the varying
distribution of dust and starlight on the line of sight.}
\label{fig:evstau}
\end{figure}

The variable geometry of dust and starlight along the line of sight
makes the relationship between color excess and extinction
non-linear, unlike the common interstellar case where an absorbing
cloud is in front of a single star.
Figure \ref{fig:evstau} plots the reddening versus the
optical depth through the disk for the Milky Way and NGC 2841 models
with no bulge dust (from Figs. \ref{fig:nobulge} and \ref{fig:n2841model})
and with scattered light included.
The projected radius along the minor axis varies along each curve, from
the near side of the disk (solid line) to the far side
(dashed line).  The same four inclination angles are shown.
The relationship between reddening and extinction is double-valued
because the near side has more reddening per unit dust column density
than the far side.

\section{Discussion}

Differential reddening between the near and far sides
of a galaxy disk may be understood as follows:  The radial
light gradients in the
bulge and disk cause the near side to have a brighter light source
{\it behind} most of the disk dust, while the same gradients
on the far side cause it to
have its brighter light source in {\it front} of the
disk dust.  Thus the disk dust has more light to block on the near side
than the far side. Consequently the near side is dimmer (Kodaira \&
Ohta 1994;
Byun, Freeman, \& Kylafis 1994)
and significantly redder. Models with the same
scale height for the dust and stars (not shown)
have even stronger colour differentials because more
of the starlight is extinguished.
Models with more disk extinction also have more differential reddening,
and models with dust in the bulge have less because
the bulge dust makes the far side redder.

Minor axis colour gradients are present
in many of the galaxies in Wray's Color Atlas (1988)
and are even more prominent in $V-K$ images of inclined galaxies
(e.g., see Peletier and Balcells 1997; Thornley
1996 \& 1997; Grosb{\o}l \& Patsis 1998; Block et al. 1994a).
Thornley (1996) found that in NGC 5055,
optical depths suddenly jump
from 0.5 on the far side, to 4-5 on the near side.

The models presented here show that such colour
gradients are a natural result of extinction from well mixed dust
embedded in
the disk, and it is predicted that for large enough inclinations,
particularly in galaxies with inner-disk dust holes, the
near-far asymmetry in V-K profiles may even be step like.
Near-side dust screens are not needed, even for the dramatic
jump of $E(V-K)\sim1$ mag in NGC 4826. There is
some sensitivity of the colour gradient to the precise distribution of
dust, such as the
presence or lack of a central hole, and this may be useful in
modelling disks.

Another point to stress is that the far side, although bluer, cannot be
considered {\it dust free} or {\it unobscured} -- terms that
continue to be used in the literature.  For example, the
Andromeda spiral M31 shows prominent dust lanes on the far side
as well as the near side
in both the IRAS 60 $\mu$m and 100$\mu$m maps (eg. Habing et al. 1984;
Walterbos \& Schwering 1987).
NGC 2841 also has arms of dust in K$'$ images
on both sides (Fig 1c in Block et al. 1996 and plate 3 in Block 1996),
as does M51 at 15$\mu$m
(Sauvage et al. 1996, Block et al. 1997).
It is simply that when the disk is well inclined to the line of sight, the
near-side dust appears to be much thicker than the
far side dust, even when both the dust and the stellar column densities
are the same on each side.

\begin{figure}
\vspace{2.6in}
\includegraphics{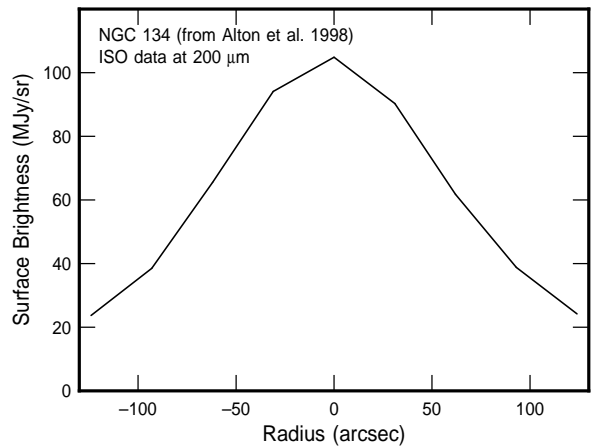}
\caption{Surface brightness along the minor axis of the galaxy NGC 134 at
200 $\mu$m, from Alton et al. (1998), showing symmetry in cold dust
emission even though this same dust presents a striking
asymmetry with foreground ``screening'' in optical images.}
\label{fig:n134}
\end{figure}

These results lead to a prediction about the
{\it emission} from cold (T$\sim20$K) galactic dust:
the far-infrared minor axis luminosity profile across a galaxy,
at wavelengths of $\sim180\;\mu$m or longer,
should be {\it symmetric}, even if the optical profile is
so asymmetic that it appears to be screened by dust on one side.
That is, optically screened galaxies will be found to have
nearly equal column densities of cold dust on both sides of the
minor axis.

A good example is NGC 134, which was cited
by Sandage \& Bedke (1994) as displaying a ``particularly
strong'' dust asymmetry between the near and far sides (see Panel 194 in
their Atlas). This galaxy has just been observed at 200$\mu$m (Alton
et al. 1998) by the Infrared Space Observatory.
The far-infrared emission profile of NGC 134, shown in figure \ref{fig:n134},
is in fact almost perfectly symmetric across the minor axis
(Alton, private communication). Thus equal amounts of cold dust,
which accounts for $\sim90$\% 
of the dust mass (Block 1996; Block et al. 1994a), reside on either
side of the minor axis, despite its apparently striking foreground dust
screen.

Another implication of the
model results, illustrated by figure \ref{fig:evstau},
is that galactic opacity is not simply related to color excess.
The conversion from color excess to opacity depends on inclination
and position in the disk.
For galaxies well inclined to the line of sight, there are many
E(V-K) values for the {\it same} optical depth, with extrema on the
minor axes and intermediate values elsewhere.
If this is not taken into account, one might erroneously
find, on the basis of E(V-K) colour excesses, that the bluer far side
contains less dust than the redder near side -- and will incorrectly
compute differences in the dust masses of the near and far sides -- although
both sides contain equal amounts.
Figure \ref{fig:evstau} does suggest, however, that the {\it average}
of the color excesses for the near and far sides might have some
simple relation to the opacity at that radius when there is no inner hole.
From the bottom panel of this figure, we obtain
$\tau\sim2.5\left(E[V-K]_{near}+E[V-K]_{far}\right)/2$.
Converting this to visual magnitudes gives
$A_V\sim2.7\left(E[V-K]_{near}+E[V-K]_{far}\right)/2$.
Note that this conversion factor
is much larger than the conventional
factor from V-K color excess to $A_V$, based on
foreground extinction for single stars, which gives
$A_V\sim 1.09 E(V-K)$ (Savage \& Mathis 1979; Whittet 1988).
The larger factor here arises because nearly
half of the dust on the line of sight is on the far side of the disk, and
this dust occults relatively few galactic stars.

Our results might also have a practical application. We have noted that
when a minor axis colour gradient is found,
the red side should be nearer.  Thus the sense of
dominant rotation in a disk galaxy can be determined from the colour gradient
if trailing spiral arms are present.  Such determinations may
be useful for crowded fields of distant galaxies in an attempt to
look for alignment or other attributes of spin orientations without
the tedious job of taking slit spectra to get rotation curves.
This characteristic of disk colour might also be useful in searching
for leading spiral arms in galaxies with rotation curves. Such arms
apparently occur in some accreting galaxies (NGC 4622: Buta et al 1992,
see also Fig 7 in Block et al. 1994b and plate 4 in Bertin
\& Lin 1996; NGC 4826: van Driel \& Buta 1993).

\section{Acknowledgements} The research of DLB is supported
by the Anglo American and de Beers Chairman's Fund Educational Trust,
and a note of great appreciation is expressed to Mrs. M. Keeton and the
Board of Trustees.

\bsp
\label{lastpage}
\end{document}